\documentclass[sigconf]{acmart}
\AtBeginDocument{%
  }

\copyrightyear{2026}
\acmYear{2026}
\setcopyright{cc}
\setcctype{by-nc-nd}
\acmConference[ICCAD '26]{IEEE/ACM International Conference on Computer-Aided Design}{November 08--12, 2026}{San Jose, CA, USA}
\acmBooktitle{IEEE/ACM International Conference on Computer-Aided Design (ICCAD '26), November 08--12, 2026, San Jose, CA, USA}
\acmDOI{10.1145/3831252.3834103}
\acmISBN{979-8-4007-2873-0/2026/11}

\usepackage{subcaption}
\usepackage{multirow}

\begin{document}

\title{Mapping Without Graphs: Learning Coherence Traffic for Task Placement}

\author{Guochu Xiong}

\affiliation{%
  \institution{Nanyang Technological University}
  \country{Singapore}
}

\author{Tianrui Ma}
\affiliation{%
 \institution{Nanyang Technological University}
 \country{Singapore}}

\author{Weichen Liu}
\authornote{
Corresponding author (email: liu@ntu.edu.sg).\\
This work is supported by the Ministry of Education, Singapore, under its Academic Research Fund Tier 2 (MOE-T2EP20224-0006).}
\affiliation{%
  \institution{Nanyang Technological University}
  \country{Singapore}}

\renewcommand{\shortauthors}{Guochu Xiong et al.}

\begin{abstract}
Cache coherence is essential for enabling communication in many-core Network-on-Chip (NoC)–based systems, yet as applications scale in size and complexity, maintaining efficient communication becomes increasingly difficult, making task mapping an effective way to mitigate these challenges. However, existing mapping approaches face two key limitations. First, existing approaches rely on predefined task graphs, which can be constructed for real applications. However, dependencies in these graphs are typically represented using abstractions such as dataflow or synchronization, derived from program structure or runtime information (e.g., traces or profiling), without explicitly capturing coherence-induced interactions arising from shared data accesses during execution. As a result, the constructed graphs provide an incomplete representation of inter-task relationships, limiting mapping effectiveness. Second, they often neglect coherence effects, even though coherence traffic is a major component of communication. This oversight creates a mismatch between design expectations and actual execution outcomes, leading to suboptimal mappings and degraded performance. To address these challenges, we propose CoTM, a coherence-aware task mapping framework that constructs a task graph by inferring inter-task dependencies from dynamic coherence behavior and applies a lightweight heuristic with a multi-start optimization strategy that iteratively refines mapping decisions using this information. A coherence-aware penalty function that integrates coherence traffic with NoC performance indicators guides the refinement process and assesses mapping quality. Experimental results demonstrate that CoTM significantly reduces link utilization by 47.85\% and lowers energy consumption by 10.30\% compared to existing approaches, underscoring the critical role of cache coherence in mapping design and highlighting its strong potential for integration with advanced NoC architectures.
\end{abstract}






\maketitle

\section{Introduction}
\label{introduction}
Cache coherence is indispensable for communication in many-core systems~\cite{b00} and is widely used across platforms ranging from smartphones to AI accelerators and high-performance computing, maintaining data consistency. Coherence protocols such as MESI~\cite{b02} ensure that cached copies are updated or invalidated, making changes visible to all cores and preventing data races, thereby achieving correct and reliable application execution. The communication required for these coherence operations is supported by the Network-on-Chip (NoC), which provides a modular and efficient interconnect fabric linking multiple processing elements (PEs) and thereby enables the transport of coherence messages such as invalidations, updates, and cache-to-cache transfers. These messages form coherence traffic that is not determined by the application alone. Instead, their volume and pattern depend on the coherence protocol, and because this traffic must be transmitted across the NoC, it increases the overall network load. As a result, coherence traffic can substantially increase NoC demand, introducing overhead that directly impacts latency~\cite{b01} and overall system efficiency.
\vspace{-10pt}
\begin{figure}[htbp]
    \centering
        \includegraphics[width=0.95\linewidth]{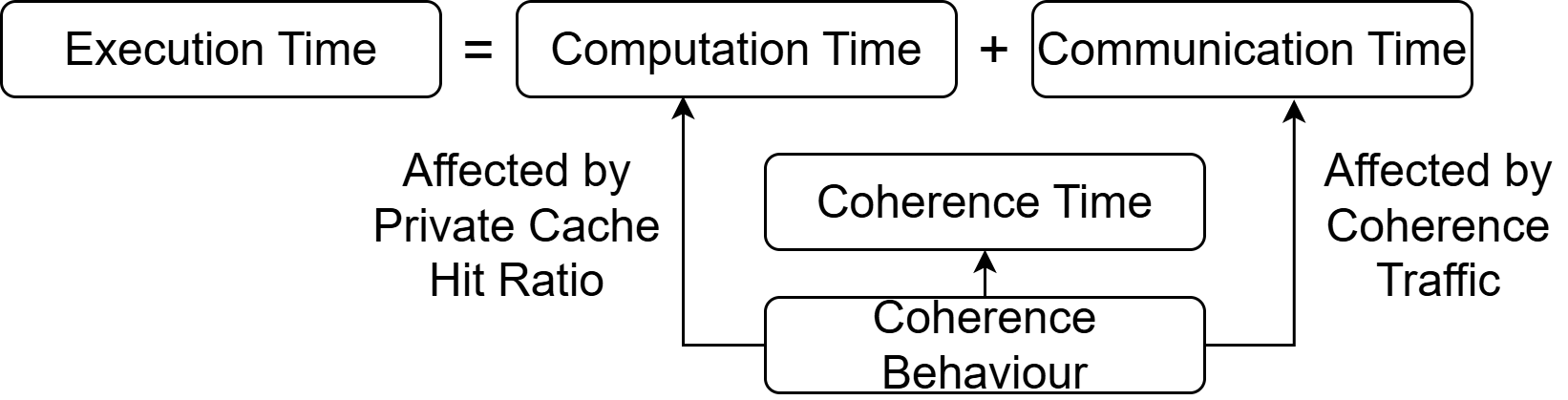}
    \vspace{-5pt} 
    \caption{Impact of coherence behaviours on task execution}
    \label{fig:coherence_impact}
\end{figure}
\vspace{-10pt}

Since applications execute as collections of interacting tasks, this overhead naturally propagates to the task level. As illustrated in Figure~\ref{fig:coherence_impact}, application execution time consists of computation and communication components, both influenced by coherence behaviors. Supported by the NoC, communication time increases when frequently interacting tasks are mapped far apart across cores, since coherence messages must traverse longer or more congested NoC paths, amplifying stalls and contention. On the computation side, when multiple tasks share heavily accessed data but reside on different cores, frequent invalidations and state transitions (e.g., cache lines bouncing between caches) reduce the private cache hit ratio, limiting opportunities for local reuse, increasing access latency, and triggering costly lookups or remote memory accesses. Collectively, these coherence-driven effects on communication and computation significantly shape execution time and ultimately determine application performance.

As a result, effective task placement is essential to mitigate these effects and improve application performance. Numerous mapping strategies~\cite{b03,b04,b05,b06,b07,b302,b303} have been proposed to address multi-dimensional challenges. However, they suffer from two key limitations.

First, most existing works focus on specific design aspects, such as NoC-related optimizations~\cite{b03}, system-level objectives~\cite{b04,b05,b06,b07} (e.g., energy efficiency and resource utilization), or cache behavior~\cite{b302,b303}, while coherence effects remain largely underexplored. Since coherence traffic can significantly increase message volume and contention under poor task placement, neglecting it creates a mismatch between design assumptions and actual execution behavior, leading to suboptimal mappings and degraded performance.

Second, existing methods typically assume predefined task graphs, where applications are modeled using graph-based representations with nodes denoting tasks and edges capturing data dependencies or communication relationships to guide the mapping process. Although such graphs can be constructed for real applications, through efforts that infer dependencies from program structure or execution behavior~\cite{b09,b18,b08,b304}, the resulting representations rely on abstractions such as dataflow or synchronization relationships derived from program semantics or runtime information (e.g., traces or profiling), without explicitly accounting for coherence-induced interactions arising from shared data accesses during execution. Since coherence traffic plays a critical role in communication during real application execution, ignoring it causes certain inter-task relationships observed at runtime to be omitted from the constructed dependency representation. Thus, two tasks may have no edge in the predefined task graph because their relationship is not captured by these abstractions, even though shared data accesses between them can still trigger coherence actions (e.g., invalidations and cache-to-cache transfers), revealing implicit data-sharing patterns that are not represented. This mismatch between modeled dependencies and actual runtime behavior leads to incomplete task graphs and suboptimal mapping decisions.

To address these limitations, it is essential to incorporate coherence effects into inter-task dependency inference for task graph construction and task mapping design, enabling a more accurate representation of application behavior and improved mapping effectiveness under realistic system conditions. In this paper, we propose CoTM, a coherence-aware task mapping framework that leverages runtime behaviors (i.e., per-core metrics observed during execution) to refine mapping decisions without relying on a predefined task graph. Our approach is motivated by two key observations: (i) Cache coherence directly shapes task-level communication and computation, making task placement a critical determinant of overall application execution performance, yet most existing mapping approaches overlook this dimension, and (ii) Existing approaches assume predefined task graphs constructed for real applications by inferring dependencies from program structure or execution behavior. However, these representations, derived from program semantics or runtime information, do not account for coherence-induced interactions arising from shared data accesses during execution. As a result, certain inter-task relationships are omitted, leading to incomplete modeling and limiting mapping effectiveness under realistic scenarios.

Building on these insights, our contributions are twofold: (i) We introduce a novel method for constructing a coherence-aware dependency graph that dynamically infers inter-task data dependencies from runtime behaviors, including both coherence and system-level metrics. This addresses the limitations of existing approaches, where task graphs, even when constructed for real applications from program structure or runtime information, rely on dependency representations that omit data-sharing patterns occurring during execution. By capturing coherence-induced interactions, our method provides a more complete representation of inter-task relationships and enables more effective task mapping. (ii) We propose a lightweight heuristic-based task mapping strategy that incorporates coherence awareness into an iterative refinement loop, augmented by a multi-start optimization mechanism. In this framework, mapping decisions are progressively improved using runtime metrics and guided by a coherence-aware penalty function that combines coherence cost with NoC communication characteristics to evaluate mapping quality. We evaluate CoTM using PARSEC benchmarks, demonstrating superior performance and efficiency compared with existing strategies.
\vspace{-10pt}

\section{Related Work}
In recent years, task mapping has gained attention for reducing communication overhead, balancing computation, and improving energy efficiency, with heuristic methods recognized for their scalability in many-core systems. To the best of our knowledge, this is the first cache-coherence-aware heuristic approach that optimizes task mapping without predefined task graphs, unifying mapping and coherence considerations in a single framework. Studies~\cite{b3,b301} comparing heuristic techniques (e.g., genetic algorithms, simulated annealing, tabu search) show that tabu-based strategies provide high-quality mappings with strong efficiency. Building on this insight, MARCO~\cite{b05} employs a tabu-based approach for dynamic workloads and reliability constraints. Mandelli et al.~\cite{b04} improve energy efficiency by packing tasks to increase utilization, while communication-aware mapping~\cite{b03} reduces contention and latency via polynomial-time placement and message prioritization. Dependency-aware methods such as HDA~\cite{b07} and HyDra~\cite{b06} further enhance locality and resource utilization by modeling inter-task interactions. Despite these advances, existing methods still rely on predefined task graphs and overlook cache coherence. Designed for non-coherent systems, they ignore the latency and congestion introduced by coherence protocols, leading to suboptimal mappings and performance loss. To partially address this, several studies explore cache behavior in mapping, such as Martin Rapp et al.~\cite{b302} and H. Ding et al.~\cite{b303}. However, the former considers only LLC latency without coherence effects, while the latter models shared-cache conflicts only for instruction accesses. As a result, these approaches miss data-sharing interactions in realistic workloads, producing mappings that fail to capture coherence-induced traffic and thus cannot represent actual execution behavior. Therefore, there remains a pressing need to (i) incorporate cache-coherence awareness into task mapping and (ii) develop graph-modeling techniques suited to realistic applications that can be effectively leveraged in mapping.

\section{Methodology}
\begin{figure}[htbp]
    \centering
        \includegraphics[width=1.02\linewidth]{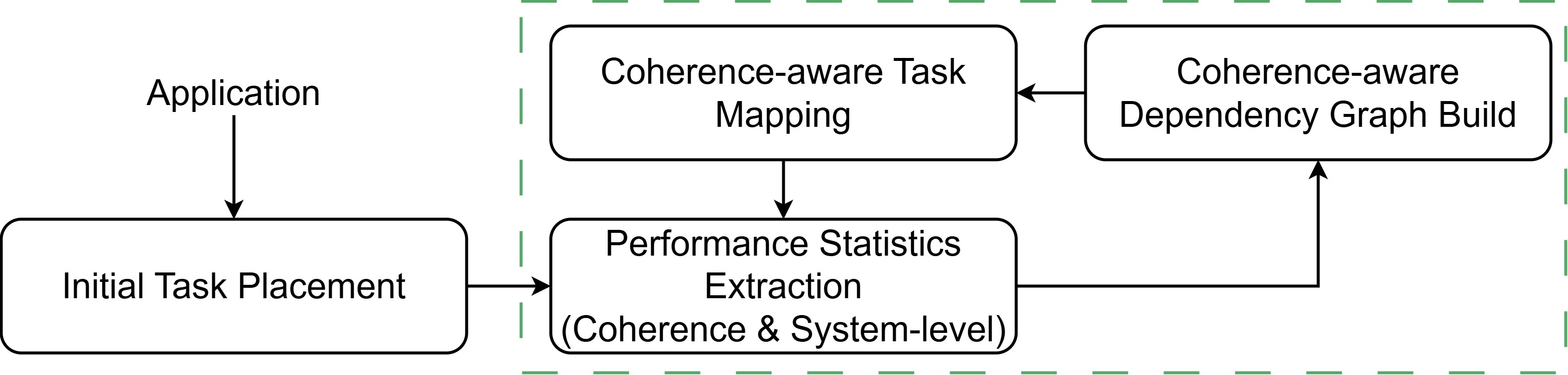}
    \vspace{-5pt} 
    \caption{Workflow of CoTM.}
    \label{fig:workflow}
\end{figure}

In this section, we present CoTM, a coherence-aware task mapping framework illustrated in Figure~\ref{fig:workflow}. CoTM operates through an iterative loop comprising two main phases: (i) constructing a task dependency graph from observed runtime behaviors, and (ii) performing lightweight heuristic-based mapping with an embedded multi-start strategy guided by this graph. In each iteration, CoTM begins with the current task-to-core mapping and runs a full-system simulation of the workload under that assignment. This execution naturally generates coherence, NoC, and system-level behaviors driven by how tasks interact through shared data. After the simulation completes, CoTM extracts per-core runtime metrics and infers inter-task relationships through a dependency score, constructing a coherence-aware task dependency graph that directs mapping refinement in the same iteration. Using this graph, CoTM constructs task mappings through a lightweight coherence-aware greedy heuristic to reduce communication overhead. To improve robustness, CoTM incorporates a multi-start strategy. Within each iteration, multiple candidate mappings are generated by varying the processing order of tasks with similar priorities and their placement choices, while preserving the priority-based scheme. These candidate mappings are evaluated using the coherence-aware penalty function, and the best-performing mapping is selected for the current iteration and used as the input for the next simulation phase. The algorithm maintains the best mapping observed across iterations and returns it as the final solution. The process continues until convergence, defined as negligible improvement in the penalty, or until a maximum number of iterations is reached.
\vspace{-10pt}
\subsection{Coherence-Aware Dependency Graph Construction}
\label{graph_build}
When targeting realistic applications, many efforts~\cite{b09,b18,b08,b304} have been made to infer dependencies from program structure or runtime information for task graph construction. However, these dependencies are typically presented through abstractions such as dataflow or synchronization relationships derived from program semantics or execution traces, while failing to capture coherence-induced interactions associated with shared data accesses during execution. Given the significant role of coherence traffic in real application communication, this limitation prevents certain inter-task relationships associated with shared data access from being reflected in the constructed graph, resulting in discrepancies between modeled dependencies and actual behavior.

In real applications, multiple tasks frequently operate on shared data. Since these tasks run on different cores, their memory accesses are coordinated through a coherence protocol. When two tasks access the same data, their cores must exchange coherence actions to maintain a consistent memory view. For example, when a core requests a cache line, the directory identifies all sharer cores and orchestrates the necessary protocol operations. The requester issues a read or write request, while sharer cores respond by supplying data through cache-to-cache transfers, invalidating their copies, or performing the required state transitions. These request–response interactions arise precisely because the tasks mapped to those cores share data, meaning coherence activity directly reflects their underlying data dependencies. Such dependencies govern execution order, data exchange, and communication overhead, making them fundamental for constructing task graphs that accurately represent real execution behavior.
\vspace{-5pt}
\begin{table}[h!]
\centering
\caption{Coherence and System Behaviour Measurements}
\vspace{-3pt}
\renewcommand{\arraystretch}{1.10}
\begin{tabular}{|p{1.3cm}|p{2.75cm}|p{3.6cm}|}
\hline
\textbf{Category} & \textbf{Feature of each core} & \textbf{Description} \\
\hline

\multirow{3}{*}{\shortstack{Coherence \\ Behaviour}}
& Coherence time
& Total delay from coherence operations includes: (i) delay from write-hit operations in the private cache when in the S state, (ii) delays from write misses, and (iii) delays from read misses.  \\
\cline{2-3}

& Num\_l1\_l1\_message
& The number of coherence-induced message transmission that a core sends to other cores' caches. \\
\cline{2-3}

& Private cache hit ratio
& The percentage of memory accesses by a core that are successfully served from its own private cache, without needing to fetch data from shared caches, other cores, or main memory. \\
\hline

\multirow{2}{*}{\shortstack{System \\ Behaviour}}
& Busy time
& Time that a core spends actively processing instructions or operations, excluding data wait time. \\
\cline{2-3}

& Execution time
& Time of completing all the tasks. \\
\hline

\end{tabular}
\label{tab:combined_measurement}
\end{table}
\vspace{-5pt}

However, although these dependencies originate from tasks, they are not visible in static code, and their corresponding coherence behaviors appear only in per-core metrics observed during execution, as shown in Table~\ref{tab:combined_measurement}. For example, a high number of coherence messages exchanged between two private caches (i.e., Num\_l1\_l1\_messages in Table~\ref{tab:combined_measurement}) indicates that the tasks mapped to those cores frequently access shared data and repeatedly trigger coherence actions. Therefore, the intensity of these activities, as reflected in the metrics, indicates how frequently tasks on different cores interact through shared memory. Consequently, tasks whose cores exhibit similarly strong coherence activity (i.e., their per-core metrics show comparable levels) are inferred to have stronger data-sharing relationships.

Based on this insight, CoTM first introduces \textbf{activity modeling} to capture the intensity of coherence-induced interactions between tasks by leveraging runtime information collected at the core level in each iteration. Each task $t_i$ is associated with the core $c_m$ assigned in the mapping result in the last iteration. Although a task executes entirely on a single core, multiple tasks may share the same core during execution, meaning that the coherence and communication behavior observed on that core reflects the aggregated activity of all tasks mapped to it rather than any individual task in isolation. To represent this aggregated behavior in a consistent and comparable manner, each core is characterized by a normalized feature vector derived from the metrics listed in Table~\ref{tab:combined_measurement}, capturing key aspects of its runtime behavior. These per-core feature vectors act as behavioral proxies for the tasks mapped to those cores in the preceding iteration, and normalization ensures that the extracted signals provide an indirect but informative estimate of each task’s coherence-related activity. Let the normalized feature vector for core $c_m$ be:
\begin{equation}
\mathbf{F}_{c_m} = \left[ f_1^{(m)}, f_2^{(m)}, \ldots, f_K^{(m)} \right]
\end{equation}
where $K$ is the number of normalized features (e.g., 5 in our implementation shown in Table ~\ref{tab:combined_measurement}), $m$ denotes the core executing task $t_i$, and $f_k^{(m)}$ represents the normalized value of the $k$-th metric (Table~\ref{tab:combined_measurement}) for core $c_m$. Using these feature representations, the activity level between two tasks $t_i$ and $t_j$, mapped to cores $c_m$ and $c_n$, is defined as:
\begin{equation}
A(t_i, t_j) = \frac{1}{2K} \left( \sum_{k=1}^{K} f_k^{(m)} + \sum_{k=1}^{K} f_k^{(n)} \right)
\end{equation}

While activity modeling captures the magnitude of coherence behavior and provides an initial indication of potential data-sharing relationships, it does not explicitly distinguish whether such activity arises from direct interactions between specific tasks. For example, two cores may both exhibit high coherence-related activity while interacting with different sets of cores, resulting in weak or no direct relationship between the corresponding tasks.

To address this limitation, CoTM further introduces \textbf{interaction modeling}, which evaluates the consistency of behavior patterns between tasks. Specifically, each core is represented by its feature vector $\mathbf{F}_{c_m}$, and interactions between tasks are inferred by comparing the feature vectors of their mapped cores. When two tasks frequently share data, the coherence protocol induces correlated effects on their corresponding cores, such as similar levels of l1\_l1\_message exchanges (Table~\ref{tab:combined_measurement}) and aligned coherence behavior. As a result, their feature vectors tend to exhibit similar patterns across multiple dimensions.

To quantify this relationship, we adopt cosine similarity~\cite{b15,b13}, which measures the alignment between feature vectors. The interaction level between tasks $t_i$ and $t_j$ is defined as:
\begin{equation}
I(t_i, t_j) =
\frac{\mathbf{F}_{c_m} \cdot \mathbf{F}_{c_n}}
{\left\| \mathbf{F}_{c_m} \right\| \, \left\| \mathbf{F}_{c_n} \right\|}
\end{equation}

Building on activity and interaction modeling, CoTM defines the coherence-induced dependency between tasks as:
\begin{equation}
D(t_i, t_j) = \alpha \cdot A(t_i, t_j) + \beta \cdot I(t_i, t_j)
\end{equation}
where $\alpha$ and $\beta$ are non-negative weights satisfying $\alpha + \beta = 1$. Therefore, tasks whose mapped cores exhibit similar levels of coherence activity and consistent interaction patterns are inferred to have stronger data-sharing relationships. By jointly capturing both the magnitude of coherence activity and the consistency of interaction patterns, the dependency score provides a unified measure of inter-task coupling. This enables CoTM to construct a coherence-aware task dependency graph that more faithfully reflects task relationships observed at runtime.

Building on this scoring mechanism, we introduce two thresholds: the strong coherence threshold $T_{\text{strong}}$ and the weak coherence threshold $T_{\text{weak}}$. Since the dependency score is defined as a weighted combination of activity level $A(t_i, t_j)$ and interaction level $I(t_i, t_j)$, both normalized to the range $[0,1]$, the resulting score also lies within $[0,1]$, providing a bounded and comparable measure of coherence-induced coupling between tasks. Under this interpretation, thresholding can be applied to distinguish different levels of task relationships. In this context, similarity-based analysis provides a useful reference for selecting threshold values. In particular, cosine similarity values around $0.7$ are commonly used to denote strong relationships~\cite{b16}. Following this convention, we set $T_{\text{strong}}=0.65$ for high-affinity coherence between tasks and $T_{\text{weak}}=0.25$ for weaker yet relevant associations. Thus, these thresholds classify dependencies into three levels: strongly coherence-dependent when $\text{D}(t_i,t_j) > T_{\text{strong}}$, moderately coherence-dependent when $T_{\text{weak}} < \text{D}(t_i,t_j) \leq T_{\text{strong}}$, and weakly coherence-dependent when $\text{D}(t_i,t_j) \leq T_{\text{weak}}$. This graded classification provides a nuanced understanding of how closely tasks interact through shared data.

To further reinforce spatial locality, we boost the dependency score by a factor $\gamma$ when both tasks are mapped to the same core ($c_m = c_n$). This reflects their ability to communicate via local caches with negligible coherence overhead. Amplifying the score in such cases encourages the preservation of high-affinity task pairs that already minimize coherence and NoC communication costs.

By integrating these components, CoTM constructs a weighted coherence-aware task dependency graph that captures the relationships among tasks based on their coherence interactions. Each node in the graph represents a task, and each edge denotes a coherence-induced relationship between two tasks. For every task pair $(t_i, t_j)$, an edge is established when their dependency score $D(t_i, t_j)$ exceeds $T_{\text{weak}}$. The corresponding dependency score is then assigned as the edge weight, reflecting the degree of inter-task coupling inferred from both activity intensity and interaction consistency. The resulting weighted graph captures not only the presence of inter-task dependencies but also their relative strength, providing a structured representation to guide subsequent mapping refinement in CoTM. To ensure robustness across diverse task granularities, CoTM adheres to three key principles: (i) behavior-driven inference, ensuring that dependency estimation reflects actual runtime coherence interactions; (ii) stability of coherence patterns, since strong coherence relationships tend to persist across task granularities, allowing consistent dependency identification; and (iii) metric normalization, which guarantees fair comparison among tasks of different sizes or loads. Together, these design principles enable CoTM to generate reliable and coherence-aware dependency graphs that accurately capture communication behavior in realistic, unstructured many-core workloads.
\vspace{-10pt}
\subsection{Coherence-Aware Task Mapping}
\label{mapping}
Building on the coherence-aware dependency graph, we propose CoTM, a coherence-aware task mapping framework. Unlike prior heuristics that rely on design-time assumptions~\cite{b06} or static models derived from predefined task graphs~\cite{b03}, CoTM captures the close relationship between task mapping and coherence (Section~\ref{introduction}), enabling adaptation to actual execution dynamics rather than fixed assumptions. During each iteration, CoTM refines the mapping using a lightweight greedy heuristic that accounts for both coherence affinity and core load. Guided by the coherence-aware dependency graph obtained in Section~\ref{graph_build}, each task is assigned a priority based on the number of neighbors with strongly coherence-dependent relationships. For a task \( t_i \), we denote its neighborhood as \( \mathcal{N}(t_i) \), and define its strong dependency set \( \mathcal{N}_s(t_i) \) as the subset of neighbors whose coherence dependency scores exceed a predefined \( T_{\text{strong}} \):
\begin{equation}
\mathcal{N}_s(t_i) = \left\{ t_j \in \mathcal{N}(t_i) \mid \text{D}(t_i, t_j) > T_{\text{strong}} \right\}
\end{equation}
Where \( \text{D}(t_i, t_j) \) represents the dependency score between tasks \( t_i \) and \( t_j \) obtained in Section~\ref{graph_build}. The task priority is then given by \( \text{Priority}(t_i) = \left| \mathcal{N}_s(t_i) \right| \). Tasks are processed greedily in descending order of priority. For each task \( t_i \), we evaluate whether it should migrate to the core of a strongly dependent neighbor \( t_j \in \mathcal{N}_s(t_i) \). Migration occurs only if the target core has remaining capacity, i.e., its task load \( L_{c_n} \) on core \( c_n \) is below an overload threshold \( L_{\text{max}} \), ensuring that excessive contention is avoided while preserving locality among strongly dependent tasks.

To evaluate mapping quality, we define a composite penalty score $\mathcal{P}$ that captures the combined impact of coherence overhead and NoC communication cost:
\begin{equation}
\mathcal{P} = \sum_{c_m \in \mathcal{C}} (\gamma_{c_m} + h_{c_m}) \cdot T_{c_m}
\label{eq:penalty}
\end{equation}
where $\mathcal{C}$ denotes the set of all cores, and $T_{c_m}$ is the number of tasks mapped to core $c_m$. The term $\gamma_{c_m}$ represents the coherence time of core $c_m$, reflecting the intensity of coherence activity, while $h_{c_m}$ denotes the average hop count, computed for each candidate mapping based on the resulting task-to-core placement and the underlying NoC topology, capturing the communication distance under the current mapping. Together, these metrics capture both the intensity of coherence interactions and the communication distance over which they propagate, enabling effective evaluation of mapping quality from both coherence and communication perspectives. This formulation serves as the objective for guiding the mapping optimization process.

Despite the structured guidance provided by the coherence-aware dependency graph, the greedy nature of the optimization makes it inherently sensitive to the initial task-to-core assignment~\cite{b14}. Early placement decisions can influence both core load distribution and subsequent migration opportunities, potentially leading to suboptimal local solutions.

To address this limitation, we introduce a multi-start optimization strategy as a key component of CoTM. Unlike conventional multi-start approaches that rely on independent random initializations~\cite{b10}, our method generates multiple candidate mappings through structured variations within the coherence-aware, graph-guided greedy mapping process. Specifically, at each iteration, tasks are prioritized based on the coherence-aware dependency graph, and diversity is introduced by varying the processing order among tasks with similar priorities as well as their placement choices, while preserving the overall priority-based scheme. This results in multiple distinct candidate mappings under the same dependency structure. The refined candidate mappings are evaluated using the composite penalty score $\mathcal{P}$ defined in Eq.~(\ref{eq:penalty}) and the best-performing mapping is selected for the current iteration. The selected mapping is then used in the next simulation phase, where runtime metrics are collected and the dependency graph is updated accordingly, enabling further refinement of the mapping. This iterative process continues until convergence, defined as negligible improvement in the penalty $\mathcal{P}$ across successive iterations, or until a maximum number of iterations is reached. Through this design, the incorporation of multiple diverse construction paths within each iteration enables CoTM to explore different regions of the solution space and reduces sensitivity to initialization. Consequently, the optimization process is guided toward higher-quality mappings, resulting in stable convergence behavior and consistently improved mapping outcomes across iterations.

By embedding coherence awareness into this iterative feedback process, CoTM provides a scalable and adaptive refinement mechanism. This design goes beyond traditional heuristic methods by ensuring that mapping decisions reflect actual execution behavior. Formally, the mapping complexity per iteration is:
\begin{equation}
T_{\text{map}} = O\big(S \cdot (E + N \log N + P)\big)
\end{equation}
where $N$ denotes the number of tasks, $P$ the number of cores, $E$ the number of edges in the dependency graph, and $S$ the number of multi-start instances. The runtime is primarily dominated by neighbor traversal $O(E)$ and task sorting $O(N \log N)$. In practice, $S$ remains a small bounded constant, and for sparse dependency graphs ($E = O(N)$), the complexity reduces to $O(N \log N + P)$. Under iterative refinement, each pass incurs this cost, resulting in a total complexity that scales linearly with the number of iterations $R$. This complexity results in lower computational overhead compared to typical metaheuristic approaches~\cite{b05,b3,b301}, such as simulated annealing or Tabu search, which typically scale with $O(R \cdot N \cdot P)$, and genetic algorithms with $O(G \cdot N \cdot P)$, where $G$ reflects the combined effect of population size and generations. Unlike these approaches, which explore large solution spaces through global or population-based search, CoTM performs structured, graph-guided refinement with a small number of multi-start instances. This design avoids combinatorial exploration while maintaining effective search capability, resulting in a lightweight and scalable mapping solution.

\section{Experiment}
Experiments are conducted using the PARSEC 2.1 benchmark suite~\cite{b11} on the Gem5 simulator~\cite{b12}, targeting realistic many-core scenarios. Energy consumption is measured with McPAT~\cite{b28}. As summarized in Table~\ref{tab:system_config}, our framework supports multiple coherence protocols under a 2D mesh topology. For evaluation, we employ a directory-based MESI protocol~\cite{b01}, widely adopted in research and industry (e.g., CHI~\cite{b120}), to ensure broad applicability and accurate coherence modeling. We conduct most experiments on a 64-core system using an $8\times 8$ mesh, except for the scalability study, which also includes 16-core and 144-core systems. All experiments use deterministic XY routing to isolate mapping effects from routing variability while accurately capturing coherence behavior.

\begin{table}[htbp]
\centering
\caption{Supporting Configuration \& Platform Parameters}
\label{tab:system_config}
\small
\vspace{-3pt}
\renewcommand{\arraystretch}{1.10}
\begin{tabular}{|p{3.1cm}|p{4.6cm}|}
\hline
\multicolumn{2}{|c|}{\textbf{Supporting Configuration}} \\ 
\hline
\textbf{Parameters} & \textbf{Specification} \\ 
\hline
Cache Coherence Protocols & Directory-based MESI/MSI/MOESI \\
\hline
Topology Type & 2D Mesh ($4\times4$, $8\times8$, and $12 \times 12$) \\
\hline
\multicolumn{2}{|c|}{\textbf{Platform Parameters}} \\
\hline
\textbf{Parameters} & \textbf{Specification} \\ 
\hline
System Architecture & NoC-based many-core systems \\
\hline
Frequency & 2 GHz \\
\hline
Flit Size & 128 bits \\
\hline
Virtual Channels per Port & 4 \\
\hline
Flow Control & Credit-based \\
\hline
Memory Hierarchy & L1D Cache (Private, 64 KB), L2 Cache (Shared, 2 MB), Memory (512 MB) \\

\hline
Cache Coherence Protocol & Two-Level Directory-based MESI \\
\hline
\end{tabular}
\end{table}

Starting from an initial task-to-core assignment (e.g., random), CoTM performs a full-system simulation to observe real execution behaviors under the current mapping. During execution, coherence-related metrics, collected using the lightweight coherence analysis tool CCTA~\cite{b17}, together with system-level indicators (Table~\ref{tab:combined_measurement}), are recorded in Gem5’s default output file (\texttt{stats.txt}). These metrics are then used for coherence-aware graph construction and mapping refinement. Although the initial mapping influences early observations, CoTM progressively refines both the dependency graph and task placement across iterations by leveraging a multi-start strategy, thereby reducing sensitivity to the initial assignment and guiding the optimization toward stable, high-quality mappings (Section~\ref{mapping}). Although CoTM collects various performance metrics after each simulation (e.g., latency and link utilization), only those listed in Table~\ref{tab:combined_measurement} are used for iterative refinement.

To evaluate CoTM’s generality under realistic workloads without predefined task graphs, we employ four PARSEC benchmarks, each partitioned into a number of tasks: Vips (36 tasks, memory-intensive image processing), Canneal (25 tasks, irregular access with high coherence pressure), Fluidanimate (16 tasks, communication-intensive scientific workload), and Blackscholes (50 tasks, compute-centric with low coherence demand). These benchmarks collectively span a broad range of coherence traffic patterns, demonstrating the applicability of our method to diverse and realistic scenarios. We compare CoTM with three representative mapping approaches: (i) Tabu-based mapping (MARCO~\cite{b05}), which applies heuristic optimization for dynamic workloads and reliability, (ii) Communication-aware mapping~\cite{b03}, which uses polynomial-time algorithms for contention-aware placement and message prioritization to reduce latency, and (iii) Dependency-aware mapping (HyDra~\cite{b06}), which performs runtime remapping to adapt to workload variations while preserving dependencies. These approaches provide a strong foundation for evaluating CoTM under realistic system conditions.

\subsection{Analysis of Graph Construction in CoTM}

\textbf{System-level Efficiency.} We first evaluate CoTM’s graph construction against a TP-PARSEC–inspired strategy~\cite{b09}. TP-PARSEC employs a code-centric preprocessing framework to perform task creation and infer task dependencies based on synchronization operations and thread-level parallelism, primarily capturing execution ordering and control relationships between tasks. To ensure fairness, we adapt TP-PARSEC to gem5 with Pthread-style parallelism so that both approaches operate under comparable task structures and runtime behavior. As shown in Table~\ref{tab:graph_comparison}, TP-PARSEC represents dependencies through synchronization and thread-level relationships, whereas CoTM constructs dependency graphs based on runtime coherence behavior, explicitly capturing data-sharing interactions between tasks. By modeling fine-grained coherence-induced interactions and filtering out weak dependencies, CoTM provides a more faithful representation of inter-task relationships and exposes safe parallelism, achieving a 15.74\% reduction in execution time compared to TP-PARSEC, while also improving applicability to real-world workloads and supporting downstream mapping designs.
\vspace{-5pt}
\begin{table}[htbp]
\centering
\caption{System-level comparison across methods}
\label{tab:graph_comparison}
\small
\vspace{-3pt}
\renewcommand{\arraystretch}{1.10}
\begin{tabular}{|p{1.9cm}|p{2.5cm}|p{3.2cm}|}
\hline
\textbf{Method} & \textbf{Dependency Type}  & \textbf{Exec. Time Improvement} \\ \hline
TP-PARSEC~\cite{b09} & Synchronization \& thread-level dependencies &$0$ (reference) \\ \hline
Our method & Coherence-induced dependencies &   $15.74\%$ \\ \hline
\end{tabular}
\end{table}
\vspace{-15pt}
\begin{figure}[htbp]
    \centering
        \includegraphics[width=0.90\linewidth]{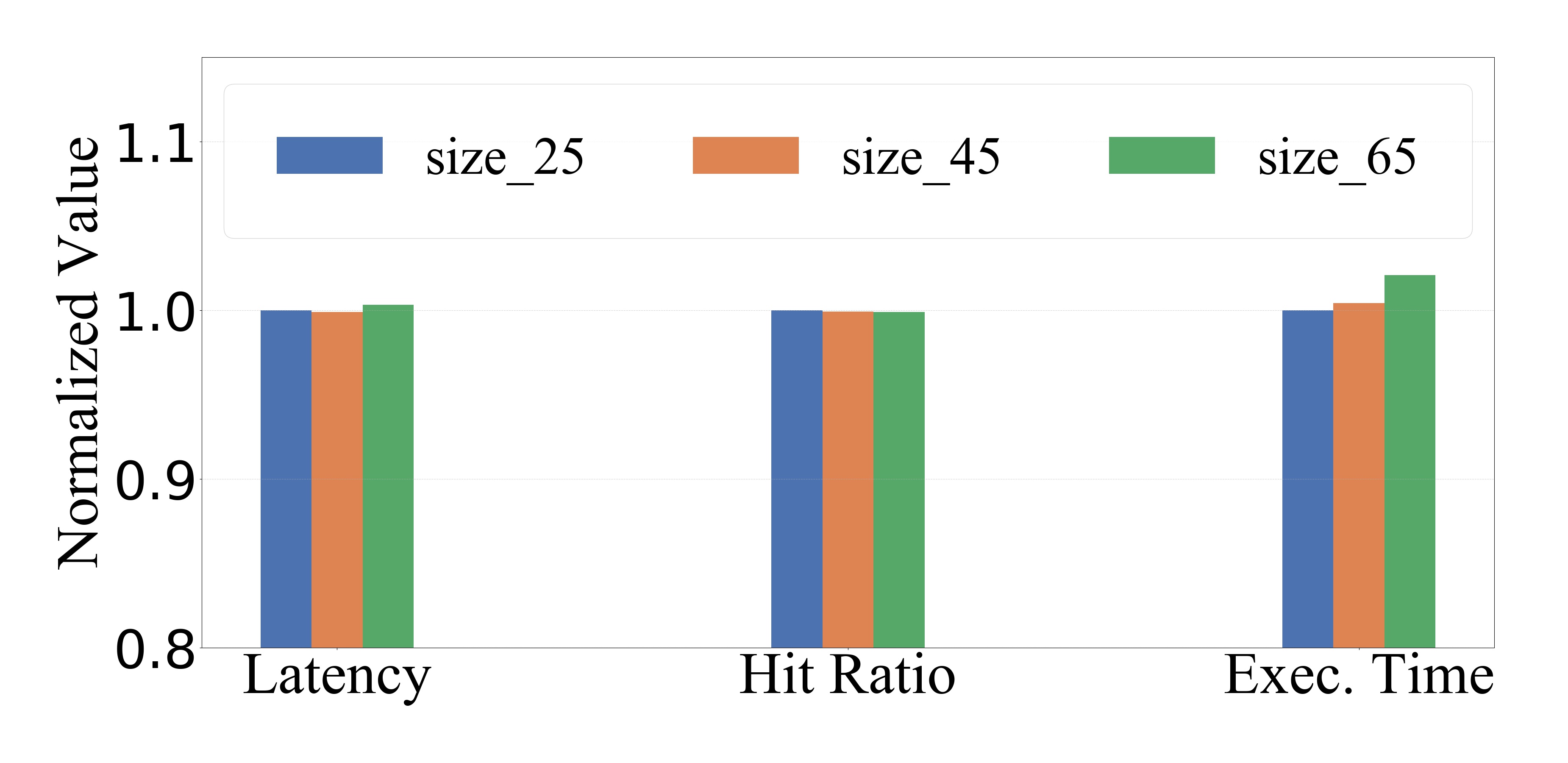}
    \vspace{-10pt} 
    \caption{Comparison of different task granularities.}
    \label{fig:task_granularity_comparison}
\end{figure}

\textbf{Tolerance on Task Granularity.} To validate robustness under varying task granularities (Section~\ref{graph_build}), we randomly partition the Canneal workload into 25, 45, and 65 tasks, representing coarse-, moderate-, and fine-grained decompositions that cover a range of design choices in task-level parallelism. As shown in Figure~\ref{fig:task_granularity_comparison}, CoTM maintains stable performance across all granularities: NoC metrics (e.g., latency) and coherence behaviors (e.g., cache hit ratio) vary within 0.3\%, while system-level metrics (e.g., execution time) deviate by less than 2\%. This robustness stems from CoTM’s ability to dynamically infer inter-task relationships and construct a coherence-aware task graph directly from runtime behaviors, rather than relying on statically defined task dependencies as in TP-PARSEC. As a result, CoTM achieves minimal performance loss regardless of how tasks are partitioned, underscoring its strong applicability and scalability across applications.

\begin{figure}[htbp]
    \centering
        \includegraphics[width=\linewidth]{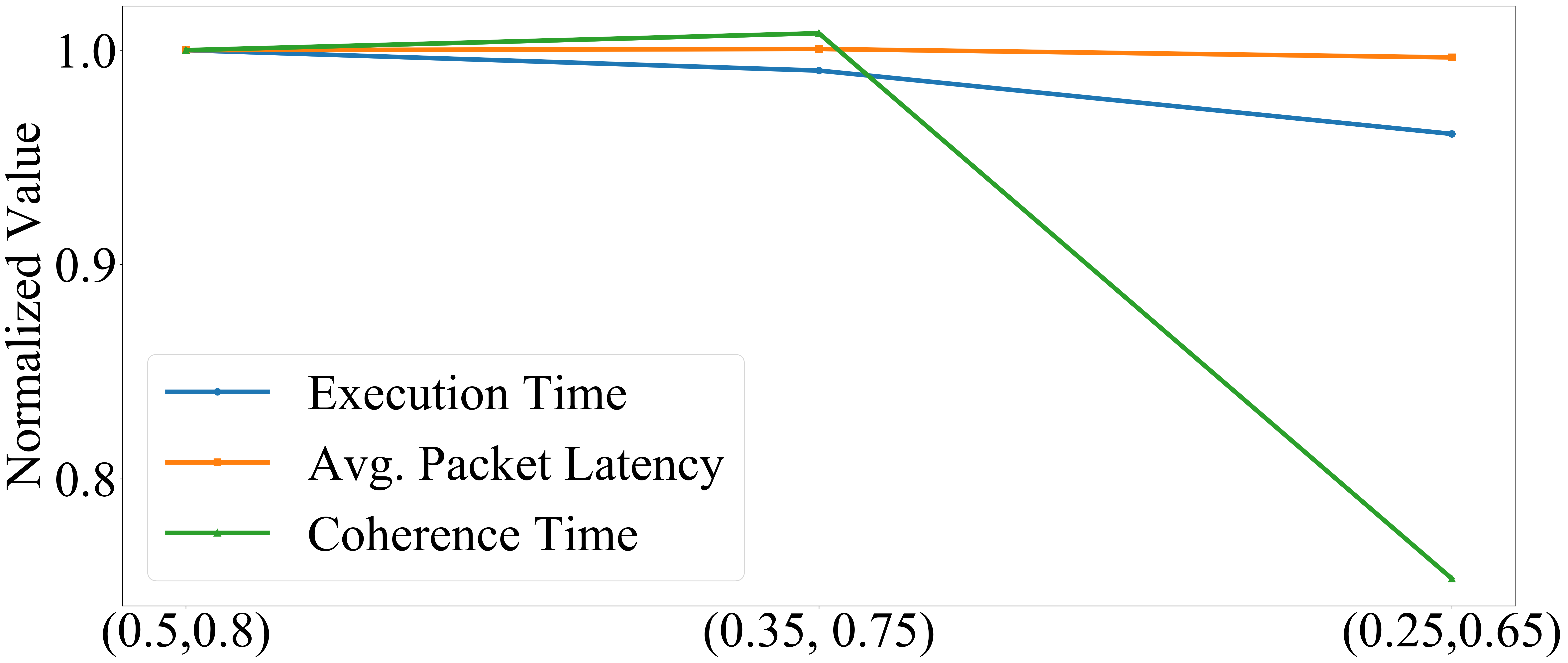}
    \vspace{-5pt} 
    \caption{Comparison of different coherence threshold pairs.}
    \label{fig:threshold}
\end{figure}

\textbf{Tolerance on Coherence Threshold.} A cosine similarity threshold of around 0.7 is commonly used to indicate strong relationships~\cite{b16} (Section~\ref{mapping}). To identify an effective coherence threshold pair, we further explore small adjustments around this value by evaluating three configurations: a moderate setting $(0.35, 0.75)$ as the reference, a relaxed setting $(0.25, 0.65)$ that admits more weak dependencies, and a strict setting $(0.50, 0.80)$ that captures only strong interactions. As shown in Figure~\ref{fig:threshold}, the relaxed setting performs best, reducing execution time by 2.98\% and 3.91\%, packet latency by 0.39\% and 0.34\%, and coherence time by 25.23\% and 24.71\% compared with the moderate and strict cases. Although NoC-level gains are modest, the substantial drop in coherence time under the relaxed threshold drives most of the execution-time improvements. These results indicate that $(0.25, 0.65)$ achieves the best trade-off by filtering out low-impact interactions while retaining those that meaningfully affect performance, highlighting that even small parameter adjustments uncovered through our further exploration can yield significant efficiency improvements.

\vspace{-10pt}

\subsection{Multi-Level Performance Analysis of CoTM}
In this section, we evaluate the multi-level performance of CoTM at the NoC, coherence, and system levels by comparing it with Tabu-based mapping~\cite{b05}, communication-aware mapping~\cite{b03}, and HyDra~\cite{b06} in a realistic many-core system.

\begin{figure}[htbp]
    \centering
        \begin{subfigure}[b]{0.45\textwidth}
        \centering
        \includegraphics[width=\linewidth]{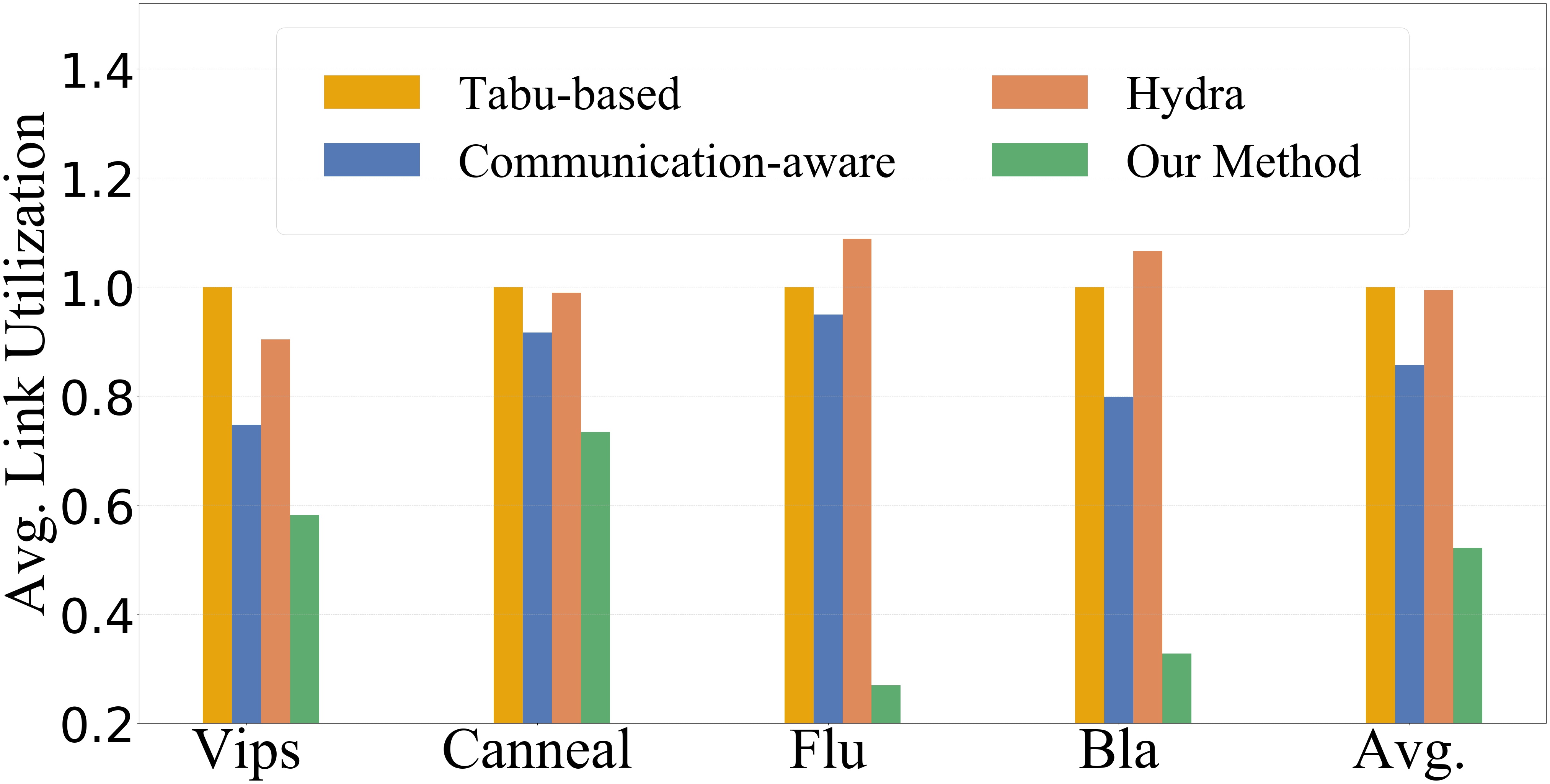}
        \caption{Average link utilization}
        \label{fig:Average_link}
    \end{subfigure}

    \begin{subfigure}[b]{0.45\textwidth}
        \centering
        \includegraphics[width=\linewidth]{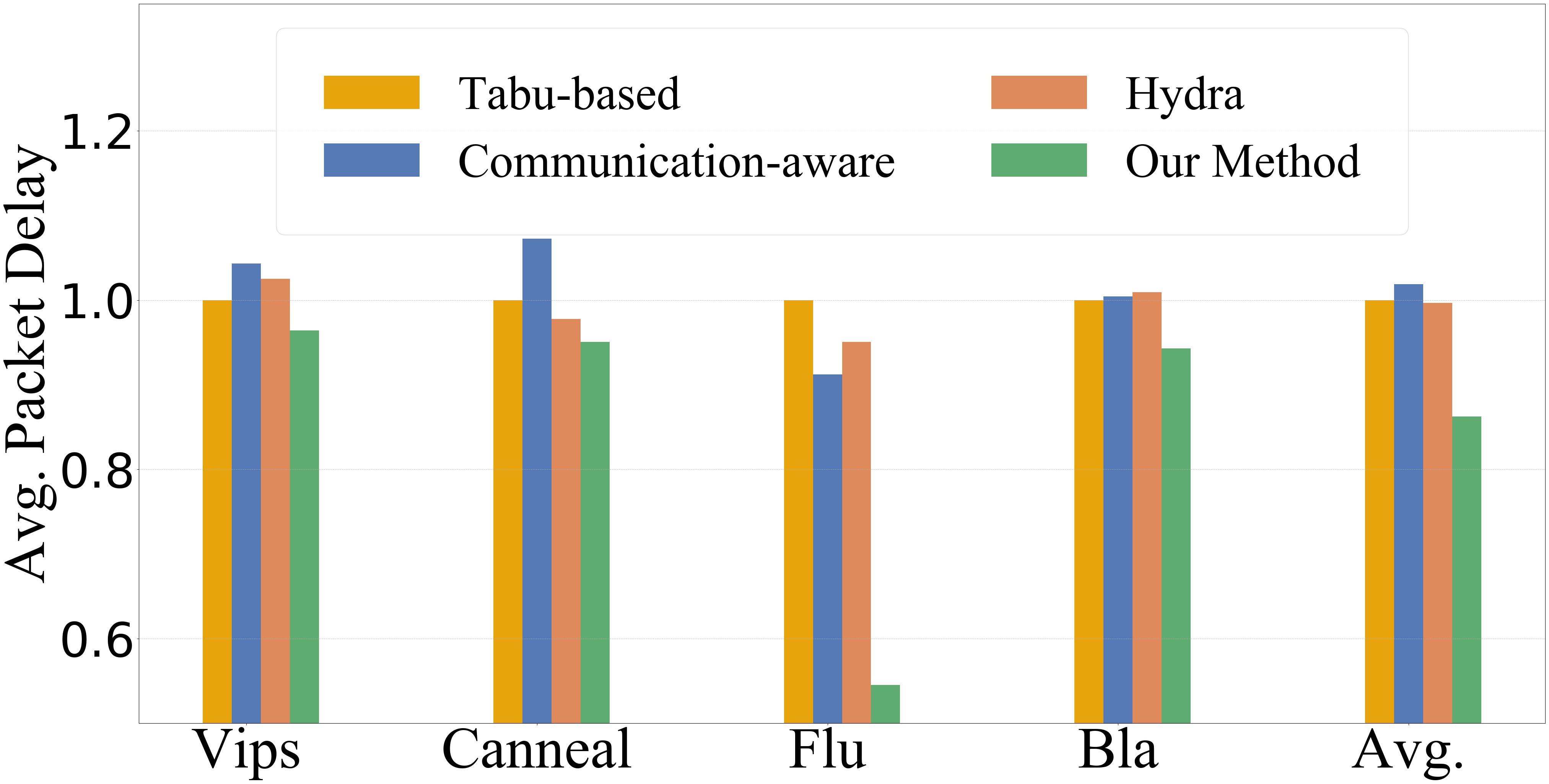}
        \caption{Average packet delay}
        \label{fig:Average_packet_delay}
    \end{subfigure}

    \begin{subfigure}[b]{0.45\textwidth}
        \centering
        \includegraphics[width=\linewidth]{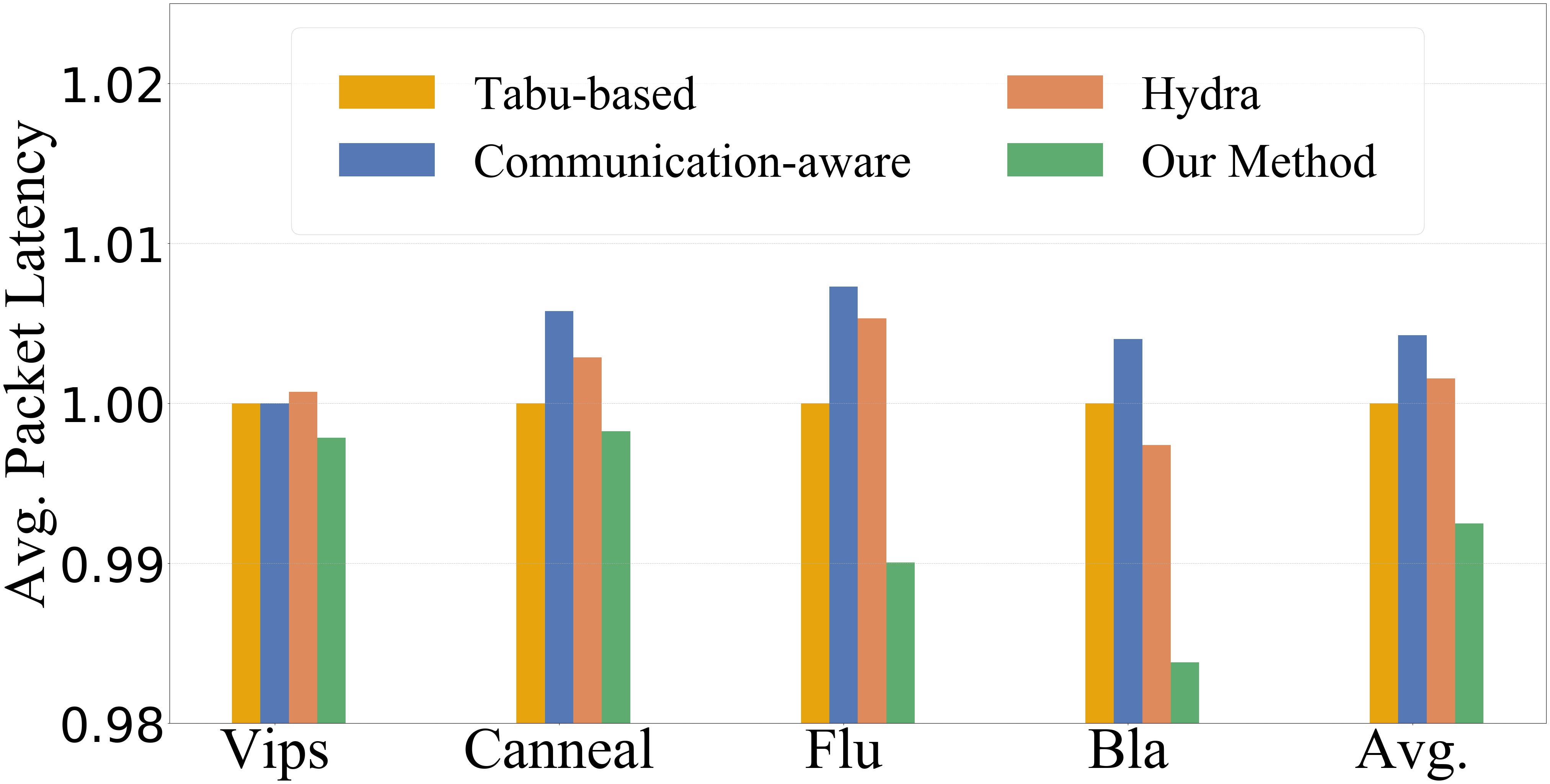}
        \caption{Average packet latency}
        \label{fig:Average_packet_latency}
    \end{subfigure}
    
    \vspace{-5pt} 
    \caption{Normalized NoC performance in PARSEC: (a) Average link utilization, (b) Average packet delay, and (c) Average packet latency.}
    \label{fig:noc_performance}
\end{figure}

\textbf{Enhancement of NoC Efficiency.}
As shown in Figure~\ref{fig:noc_performance}, CoTM reduces average link utilization by 47.85\%, 39.16\%, and 47.55\%, and average packet delay by 13.76\%, 15.33\%, and 13.47\% compared to Tabu-based, communication-aware, and HyDra methods, respectively, while keeping average packet latency within 2\%.

The reduction in link utilization and packet delay is primarily attributed to coherence-aware task placement, which co-locates coherence-intensive tasks on nearby cores. This localizes coherence traffic and reduces the number of links involved in data transfer, thereby lowering overall network load and reducing contention at routers. Consequently, packets experience shorter waiting times, leading to reduced queueing delay and lower packet delay. In contrast, although Tabu-based mapping optimizes placement based on global communication cost, communication-aware mapping focuses on contention-aware placement, and HyDra performs runtime remapping to adapt to workload variations, none of these approaches account for coherence-induced interactions, which play a critical role in realistic communication behavior. As a result, coherence-related traffic tends to remain more dispersed across the network, sustaining higher contention and limiting the achievable reduction in delay.

This effect is more pronounced in communication-intensive workloads such as fluidanimate (flu), where frequent data exchanges generate sustained network pressure. In such cases, CoTM effectively captures coherence-induced interactions and localizes the resulting traffic, significantly reducing contention. This also leads to lower link utilization, as heavily used communication paths are replaced by shorter, localized transfers. In contrast, the improvement is less pronounced for other workloads due to their distinct communication characteristics. For memory-intensive workloads with relatively structured access patterns, such as vips, communication is already partially localized under existing mappings, leaving limited room for further improvement in both delay and link utilization. For applications with irregular access patterns, such as canneal, communication tends to be more dispersed and less predictable, reducing the effectiveness of traffic localization and thus limiting gains. Finally, for workloads with low coherence demand, such as blackscholes (bla), inter-core data sharing is minimal, resulting in low coherence traffic and reduced network injection. In this case, CoTM further concentrates the already limited communication, leading to a noticeable reduction in link utilization, while the impact on delay remains modest due to the inherently low network pressure.
Despite reductions in link utilization and packet delay, improvements in packet latency remain modest. This is because latency is influenced not only by congestion reduction but also by factors such as router pipeline delays and bursty traffic, which are less sensitive to task mapping. While reduced congestion lowers average queueing delay, fixed per-hop costs and transient traffic fluctuations can still introduce localized queuing and limit overall gains. Furthermore, some inter-task communications inevitably span longer distances due to task dependencies and resource constraints, contributing latency that cannot be eliminated through task placement alone. As a result, even with reduced network load, overall latency improvements remain limited.

Overall, these results demonstrate that CoTM is effective under realistic communication behavior, where coherence-induced interactions play a critical role. By capturing these interactions, CoTM improves traffic locality and reduces both contention and link utilization, while consistently delivering performance benefits across diverse workloads. Moreover, it provides additional opportunities for integration with other NoC design techniques, such as routing, to further enhance system performance.

\begin{figure}[htbp]
    \centering
        \includegraphics[width=0.95\linewidth]{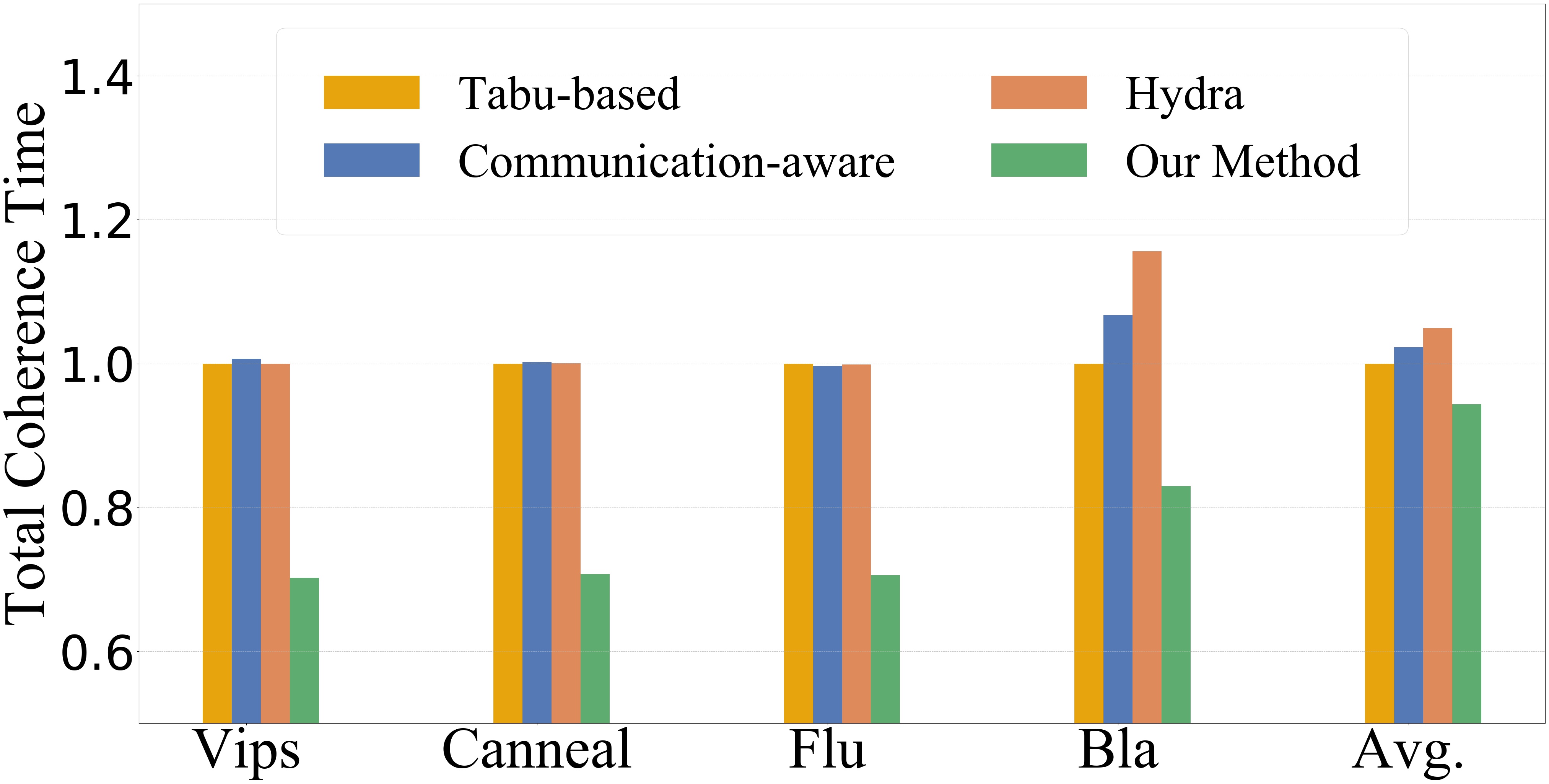}
    \vspace{-5pt} 
    \caption{Normalized coherence time in PARSEC.}
    \label{fig:cc_time}
\end{figure}

\textbf{Enhancement of Coherence Performance.}
As shown in Figure~\ref{fig:cc_time}, our method reduces coherence time by 5.63\%, 7.70\%, and 10.06\% compared to Tabu-based, communication-aware, and HyDra mappings, respectively. These improvements stem from coherence-aware task placement guided by runtime metrics. By jointly considering activity intensity and interaction consistency, CoTM identifies strongly coupled tasks and relocates them to nearby cores, improving spatial locality, reducing inter-core interactions, and shortening communication paths. As a result, coherence-related communication becomes more localized, reducing communication distance and contention, thereby lowering the overall cost of coherence operations and reducing coherence time.

\begin{figure}[htbp]
    \centering
        \begin{subfigure}[b]{0.44\textwidth}
        \centering
        \includegraphics[width=\linewidth]{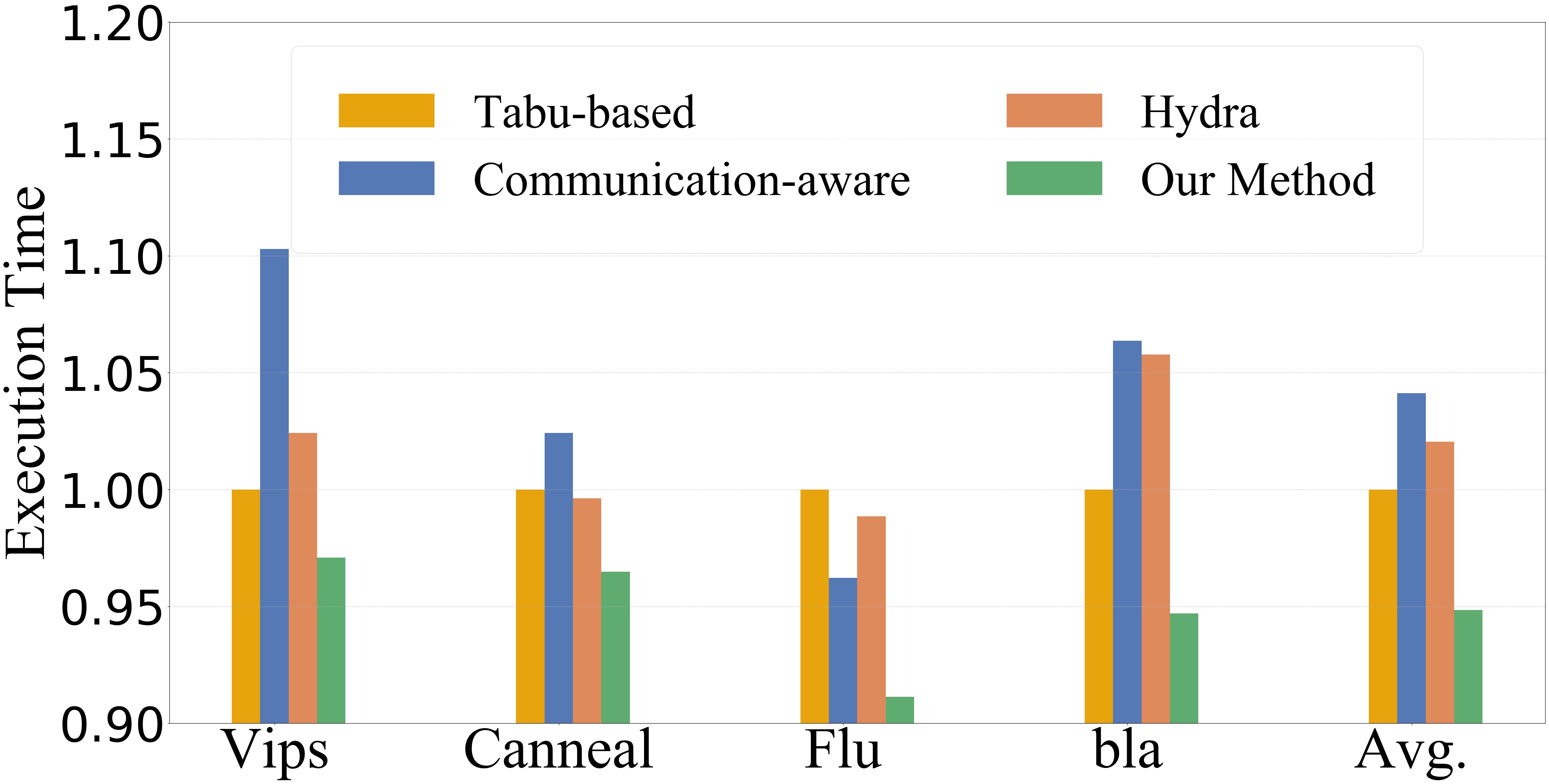}
        \caption{Execution time}
        \label{fig:exe_time}
    \end{subfigure}

    \begin{subfigure}[b]{0.44\textwidth}
        \centering
        \includegraphics[width=\linewidth]{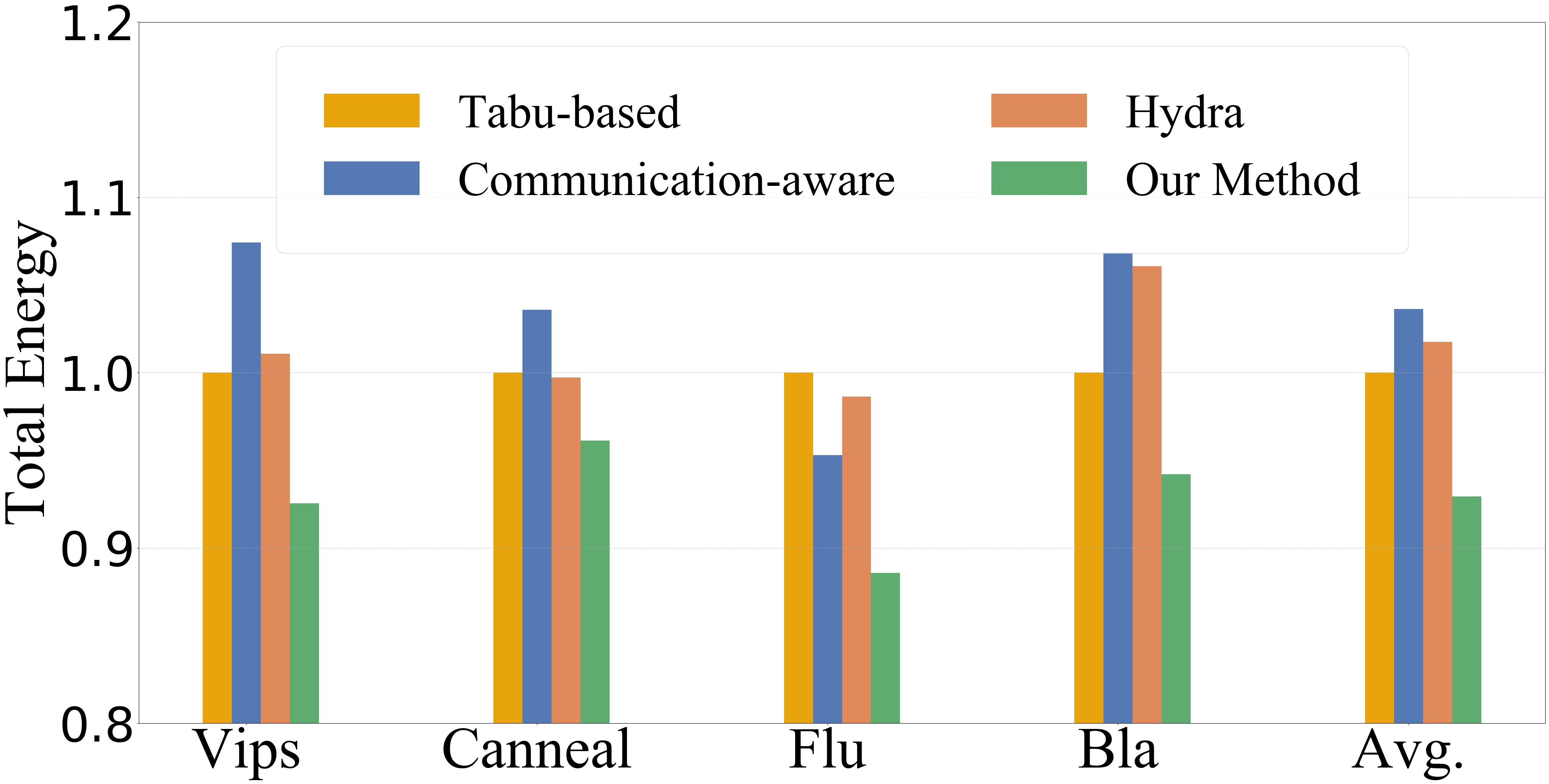}
        \caption{Total energy}
        \label{fig:energy}
    \end{subfigure}

    \vspace{-5pt} 
    \caption{Normalized system-level performance in PARSEC: (a) Execution time, and (b) Total energy.}
    \label{fig:sys_performance}
\end{figure}

\textbf{Enhancement of System Efficiency.}
As shown in Figure~\ref{fig:sys_performance}, CoTM reduces total execution time by 5.15\%, 8.91\%, and 7.06\%, and total energy consumption by 7.06\%, 10.30\%, and 8.66\% compared to Tabu-based mapping, communication-aware mapping, and HyDra, respectively. By leveraging runtime metrics, CoTM reduces coherence time and mitigates delays caused by cache-to-cache interactions and memory stalls. Meanwhile, improved NoC utilization alleviates network congestion. Together, these effects contribute to faster application execution. The reduction in energy consumption arises from both shorter execution time and improved communication efficiency. By localizing coherence-related traffic, CoTM minimizes unnecessary data movement and reduces activity across links and routers, thereby lowering dynamic energy consumption in the NoC. In addition, fewer coherence-induced stalls shorten execution time, reducing the duration of both computation and communication, which further enhances overall energy savings.

We further validate scalability by evaluating CoTM on $4\times4$ and $12\times12$ mesh systems. In both cases, CoTM delivers execution-time improvements consistent with the 64-core results, with deviations within 0.1\%. Convergence varies by workload: For instance, on the $4\times4$ system, Vips converges in about 15 iterations, yielding roughly a 12\% execution-time improvement, while Canneal converges in around 6 iterations with an improvement of about 8\%. These results show that CoTM adapts effectively to diverse workloads while maintaining consistent benefits across system sizes.

\section{Conclusion}
This work shows that coherence behavior is a critical determinant of application execution in many-core systems. Existing mapping approaches rely on task graphs whose dependencies are derived from program structure or runtime abstractions, failing to capture coherence-induced interactions arising from shared data accesses during execution. This leads to incomplete dependency representations and limits mapping effectiveness. Moreover, by overlooking cache coherence dynamics, these approaches create a mismatch between design assumptions and actual execution behavior, resulting in suboptimal mappings and degraded performance. To address these limitations, we propose CoTM, a coherence-aware, heuristic-based framework that constructs dependency graphs from runtime coherence behavior and iteratively refines task placement using a coherence-aware penalty function. Experimental results show that CoTM reduces link utilization by 47.85\% and energy consumption by 10.30\%, highlighting the importance of incorporating coherence into task mapping and its potential for integration with advanced NoC designs.

\end{document}